\documentclass[12pt]{iopart}
\usepackage{setstack}
\usepackage{iopams}
\usepackage{amssymb}
\usepackage{array}
\usepackage{multirow}

\usepackage{natbib,har2nat}

\usepackage[pdfstartview=FitH,bookmarksopen=false,colorlinks,citecolor=blue]{hyperref}
\usepackage{graphicx}

\usepackage[usenames]{color}
\usepackage{xspace}
\newcommand{\eqnref}[1]{(\ref{#1})}
\newcommand{\figref}[1]{Figure \ref{#1}}

\newcommand{\tabref}[1]{Table \ref{#1}}

\newcommand{\argmin}{\mathop{\mathrm{argmin}}}
\newcommand{\argmax}{\mathop{\mathrm{argmax}}}
\newcommand{\pixel}{\mathbf{x}}

\begin{document}

\title[Voxel-Based Dose Prediction with Multi-Patient Atlas Selection]{Voxel-Based Dose Prediction with Multi-Patient Atlas Selection for Automated Radiotherapy Treatment Planning} 
\author{Chris~McIntosh$^1$, Thomas~G.~Purdie$^{1,2}$}
\address{$^1$ Department of Medical Imaging \& Physics, Princess Margaret Cancer Centre, University Health Network (UHN), Toronto, ON, Canada}
\address{$^2$ Department of Radiation Oncology, University of Toronto, Toronto, ON, Canada}
\ead{\mailto{chris.mcintosh@rmp.uhn.on.ca},\mailto{tom.purdie@rmp.uhn.on.ca}}

\date{\today}

\begin{abstract}
Automating the radiotherapy treatment planning process is a technically challenging problem. The majority of automated approaches have focused on customizing and inferring dose volume objectives to used in plan optimization. In this work we outline a multi-patient atlas-based dose prediction approach that learns to predict the dose-per-voxel for a novel patient directly from the computed tomography (CT) planning scan without the requirement of specifying any objectives. Our method learns to automatically select the most effective atlases for a novel patient, and then map the dose from those atlases onto the novel patient. We extend our previous work to include a conditional random field for the optimization of a joint distribution prior that matches the complementary goals of an accurately spatially distributed dose distribution while still adhering to the desired dose volume histograms. The resulting distribution can then be used for inverse-planning with a new spatial dose objective, or to create typical dose volume objectives for the canonical optimization pipeline. We investigated six treatment sites (633 patients for training and 113 patients for testing) and evaluated the mean absolute difference (MAD) in all DVHs for the clinical and predicted dose distribution. The results on average are favorable in comparison to our previous approach (1.91 vs 2.57). Comparing our method with and without atlas-selection further validates that atlas-selection improved dose prediction on average in Whole Breast (0.64 vs 1.59), Prostate (2.13 vs 4.07), and Rectum (1.46 vs 3.29) while it is less important in Breast Cavity (0.79 vs 0.92) and Lung (1.33 vs 1.27) for which there is high conformity and minimal dose shaping. In CNS Brain, atlas-selection has the potential to be impactful (3.65 vs 5.09), but selecting the ideal atlas is the most challenging. 

\end{abstract}

\maketitle 

\section{Introduction}
\label{sec::Intro}

The delivery of radiotherapy (RT) is a complicated process that requires both clinical and technical expertise. Manual RT planning often involves multiple planning iterations, creating and updating dose-volume objectives and the creation of regions of interest (ROIs) solely to facilitate planning. The planning process therefore necessitates tremendous human resources often requiring hours to days to plan each patient \cite{das2009analysis}. There is also considerable variability in planning requirements depending on the treatment site and treatment technique, which leads to high inter- and intra-institutional variation in clinical practice \cite{ohri2013radiotherapy,nelms2012variation,nelms2012variationsb}. In addition, plan quality has been positively correlated with planning experience and the time invested in an individual plan \cite{batumalai2013important}. The variability in expertise and time, combined with the complex RT process. As a result, sub-optimal RT plans are used clinically; several studies have shown that RT plans that deviate from established clinical guidelines result in worse patient outcomes \cite{peters2010critical,abrams2012failure}.

There has been interest to overcome these challenges by incorporating automation into the conventional treatment planning process. These planning pipelines have traditionally incorporated historical treatment planning data with algorithms to estimate dose-volume objectives based on a limited number of features. \cite{appenzoller2012predicting,wu2013using,yang2013overlap}. In \cite{mcIntosh2016TMIAutoPlan} we introduced a novel planning pipeline that predicts a probabilistic estimate of the dose at each voxel in the image volume, thus creating a spatial dose-volume objective (SDO). The SDO punishes deviation from a specified dose value at each voxel, instead of trying to achieve a fixed amount to a volume. Our approach uses machine learning and radiomic image features to estimate the dose-per-voxel based on the dose to voxel-feature relationship observed in the most similar patients from a training database. Further, we learn to automatically select the most similar patients based on the most relevant features to dose prediction scored using the gamma metric \cite{low2003evaluation}. We incorporate features at various data scales and learnable contextual information about image appearance from the raw image volume. We aim to accurately match a novel patient to the patients in a gold standard training database with the most similar geometrical anatomy from both contoured and non-contoured structures. We output a probabilistic estimate of dose-per-voxel, enabling future development of new dose objectives. For simplicity, we refer to this as the probabilistic dose distribution. From the probabilistic distribution we previously generated a scalar dose distribution using a maximum-a-posterior estimate, which treats neighboring dose voxels independently and thus cannot constrain the dose to a desired DVH.

Our main contributions in this work are: to extend the method in \cite{mcIntosh2016TMIAutoPlan} to use a hybrid atlas-learning phase incorporating features based on the spatial dose distribution scored using the gamma metric \cite{low2003evaluation} and features derived from DVHs in conjunction with a joint dose prior; and, to extend the previous validation to three novel clinical sites: CNS Brain, Lung, and Rectum. Specifically, we introduce a conditional random field (CRF) model \cite{Lafferty2001} to transform the probabilistic distribution of dose into a scalar dose distribution that adheres to a predicted joint probability function prior (analogous to a DVH) of dose values for the target and pertinent organs-at-risk (OARs). As a result, this is the first work to balance the complementary goals of generating dose distributions that are spatially appropriate over the entire dose grid while still ensuring the proper distribution of dose within delineated ROIs is achieved.

Our hypothesis is that multiple-atlas-based dose prediction with CRF-optimized dose priors can accurately predict a resulting dose distribution without having to specify any objectives. The method can provide the required dose prediction component for a fully automated planning solution and can readily be applied to multiple clinical sites and treatment modalities using a single framework.

\section{Methods and Materials}
\label{sec::methods}

\par Automated dose prediction training, testing and analysis was conducted using patients from six treatment sites (\tabref{tb::dataSummary}). Patients were consecutively selected for each treatment site, with all patients clinically treated at our institution between 2011 and 2014. All patients were part of a retrospectively approved institutional research ethics board approval. Patients were simulated, immobilized and treated consistently within each treatment site according to the respective site-specific protocol \tabref{tb::dataSummary}.

\begin{table}
\caption{Summary of clinical data for each treatment site.}
	\label{tb::dataSummary}
\footnotesize
\lineup
\begin{indented}
\item[]\begin{tabular}{lllllll}
\br
\textbf{} & \textbf{Total} & \textbf{} & \textbf{}& \centre{3}{\underline{\textbf{\hspace{45pt}Patients\hspace{45pt}}}}\\
\textbf{Site} & \textbf{Dose} & \textbf{Fractions} & \textbf{Technique} & \textbf{Total} & \textbf{Training}	& \textbf{Testing}\\
\mr
Breast Cavity & 1000 cGy & 5 & IMRT & 117 & 97 & 20 \\
Whole Breast & 4240 cGy & 16 & IMRT & 163 & 144 & 19 \\
CNS Brain & 6000 cGy & 30 & IMRT & 130 & 113 & 17 \\
Prostate & 7800 cGy & 39 & VMAT & 164 & 144 & 20 \\
Lung & 4800 cGy & 4 & VMAT & 94 & 77 & 17 \\
Rectum & 5000 cGy & 25 & VMAT & 78 & 58 & 20 \\ \ns\ns\ns
&&&&\crule{3}\\
&&&Total:&746&633&113\\
\br
\end{tabular}
\newline
\begin{tabular}{@{}ll}

CNS: & Central Nervous System \\
IMRT: & Intensity Modulated Radiation Therapy \\
VMAT: & Volumetric Modulated Arc Therapy \\
\end{tabular}
\end{indented}
\end{table}

\begin{figure*}
\label{fig::flowChart}
	\begin{center}
		\begin{tabular}{c}
        \includegraphics[width=\linewidth,clip=]{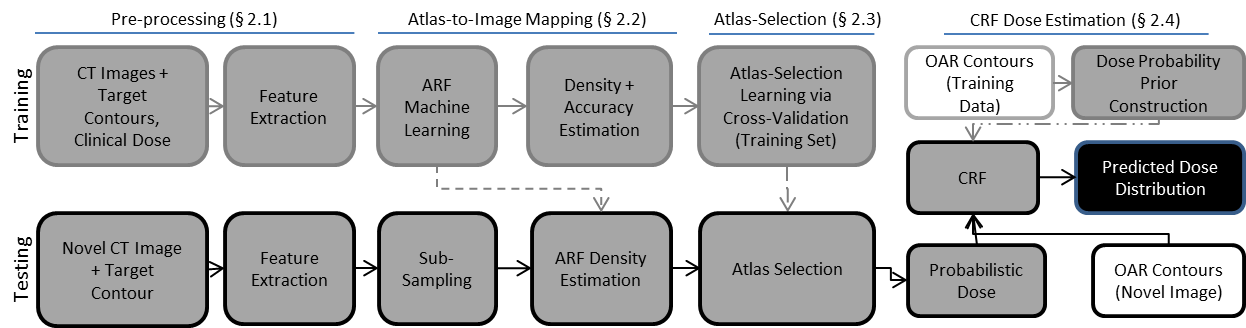}\\
		\end{tabular}
	\end{center}
	\caption[]{Flow chart showing training and testing pipelines of proposed voxel-based dose prediction algorithm. Dashed lines indicate learned output models from the training phase being used to predict dose for novel images.} \end{figure*}

An overview of the proposed process is presented in \figref{fig::flowChart}. For simplicity, since all of our image processing is performed in 3D we refer to a 3D CT image volume simply as an image, and specify a particular slice where appropriate. Training images are first loaded and characterized into image features. The features describe the appearance and texture of each voxel across a variety of scales. Machine learning using atlas regression forests (ARFs) is used to determine which features are relevant to dose prediction. Statistics over the features are used to characterize each complete image (density estimation). The ARFs are then cross-validated across all pairs within the training set, and the dose prediction accuracy of each ARF on all other training images is estimated. A second machine learning step is trained to predict the accuracy of a given ARF for each patient training image using the learned image descriptor (density estimate) as input. A planning image for a novel patient is first characterized into features, which are used to compute a density estimate for the training ARFs. The density estimate is used to predict the accuracy of each ARF on the novel patient image. The ARFs with the highest predicted accuracy are then used to predict dose for the novel patient image. We predict probabilistic dose estimates at each image voxel, i.e. a probability distribution function (PDF) over the range of potential dose at each voxel. In this work we introduced a joint voxel optimization process through a CRF to find the most likely spatial distribution of voxel under an inferred joint probability dose prior from the most similar atlases.

Complete details of the contextual atlas regression forest planning pipeline are presented in \cite{mcIntosh2016TMIAutoPlan}. In what follows we provide a higher-level summary in an effort to keep this manuscript more self-contained. For ease of reference the mathematical notation is kept consistent. There are four main components: feature extraction; atlas-to-image mapping; atlas-selection; and CRF dose prediction.

\subsection{Pre-Processing}

The data outlined in \tabref{tb::dataSummary} is divided into two independent sets per treatment site. The training data consists of patient CT planning images, contours of the targets and relevant OARs (\figref{fig::dvhErrorsAllROIs}), and corresponding clinical dose distribution pairs.

We begin by defining our notation. Let $P_j$ be an individual RT plan from a set of $M$ patients and plans $\mathbf{P}$. Each plan contains a CT image, a specification of dose-per-voxel, and one or more ROIs. The image is loaded from a DICOM CT image volume, $I_j(\pixel)$ where $\pixel\in\Omega$, the image domain, with $|\Omega|$ voxels. The set of ROIs is denoted as $\{C_{1,j}...C_{k_j,j}\}$ for plan $P_j$ with $k_j\in{\left[1,\infty\right)}$ ROIs. For simplicity, $C_{1,j}$ will generically refer to the target ROI for a particular plan, and multiple targets will appear sequentially in the set. Finally, the dose distribution is specified as $d_{j,\pixel}$, though we will equivalently write $d_j(\pixel)$ where it improves clarity.

We define a set of features, $\mathbf{F}$, with $N$ features per voxel. An individual feature $\{F_{h,j,\pixel}:h \leq  N\}$ is calculated from plan $P_j$, and voxel $\pixel$. Rather than always writing $1....n$ for some set with $n$ elements, we will use $*$ to denote taking the entire set over a particular index, or group of indices. Therefore, $F_{*,j,\pixel}$ is the same as $\{F_{1,j,\pixel}...F_{N,1,\pixel}\}$. For ease of reference our notation is summarized in \tabref{tb:notation}. 

\begin{table}
	\caption[Summary of notation.]{Summary of notation.}
	\label{tb:notation}
    \footnotesize
	\begin{indented}
		\item[]\begin{tabular}{l|l} 
        			\br
			\textbf{Notation} & \textbf{Meaning}\\
            \mr
			$\mathbf{P}$ & Set of plans with image-dose pairs: $I_j$ and $d_j$\\
			$N$ & Number of features per voxel\\
			$F_{*,j,\pixel}$ & Features $1...n$ from pixel $\pixel$ from image $I_j(\pixel)$\\
			$F_{*,j,*}$ & All features from all pixels in image $I_j(\pixel)$\\
			$P(d_{j,\pixel}|F_{*,j,\pixel})$ & Probability of dose at $\pixel$ given features $1...n$\\
			$\mathbf{T}_j$ & Atlas regression forest trained from plan $P_j$\\
			\br
		\end{tabular}
	\end{indented}
\end{table}

\subsubsection{Feature Extraction:} Following \cite{mcIntosh2016TMIAutoPlan}, feature extraction uses both non-contoured and contoured image data. First, the patient external is extracted from the image via thresholding and morphological operations, or obtained from an existing external ROI where available. In order to capture features across a variety of scales we perform convolution with a 3D texture filter bank that is an extension of the Leung-Malik filter bank \cite{leung2001}. For every scale and orientation we use one first and one second derivative of an anisotropic Gaussian filter parameterized by the scale $\sigma$ at $\sigma_x=\sigma,\sigma_y=3\sigma,\sigma_z=\sigma/3$ where use a set of scales with $\sigma=\{24,48,64\}$. For each scale, the set of orientations and rotations is parameterized by the azimuth, zenith, and angular rotation about the vector in a spherical co-ordinate system with $x,y,z$ aligned to the in-slice imaging plane from the CT image. We focus on the in-slice imaging plane (e.g. axial) with 6 samples, and 4 samples in each of the remaining orthogonal planes. This leads to $6+4+4=14$ filter orientations per scale and filter type (i.e. first or second derivative), for a total of $3\times14\times2=84$ rotational filters. We also use rotationally invariant filters in the form of isotropic Gaussians and Laplacian of Gaussians taken at set scales. The rotationally invariant filters are taken at a scale of $10$, leading to a total of $86$ filters. Filtering is performed in millimetres in world co-ordinates to account for variable voxel spacing during RT planning image acquisition. Filter parameters were selected through manual iteration on the training data and remain fixed for all experiments. Increased sampling in both scale and orientation space was added until no impressionable change was observed, with selection of relevant features left to the regression forest algorithm. Finally we include 4 target specific features for each target: a signed distance transform of the target $\Phi_{\pixel}$, and a vector in 3D denoting the direction and distance to the closest point on the target boundary:
$\left(x-c\mid c = \argmin_{c} \parallel x-c {\parallel}_2 \forall c \in \mathbf{C_1}\right)$, where $\mathbf{C_1}$ is the target.

\subsection{Atlas-to-image mapping}

The first phase of learning defines an image feature set for each pair. Using the characterized features, an ARF is learned for each pair that models the relationship between a patient's image features and their clinical dose plan on a voxel-by-voxel basis. Each ARF also computes a  probability density function estimate to measure the probability of observing the image features given the training image. For example, the density estimate learns that a particular training image having a target near a given OAR (e.g. in Rectum, the dose distribution will pinch in the region of the PTV when the Small Bowel is in close proximity)  was a key feature in generating its dose distribution. The next learning phase will then leverage these density estimates to characterize the image as a whole, and learn to estimate which images will have their corresponding dose distributions accurately predicted by a given ARF (i.e. perform atlas-selection).

First introduced by Leo Breiman in 2001, random forests (RFs) are a generalization of decision trees that use a mode-based voting algorithm over a set or ‘forest’ of decision trees \cite{Breiman2001}. A novel sample is classified by each decision tree, and then the mode of the output over the entire forest is taken as the final output class. RFs are a non-parametric regression algorithm where a regression tree, estimating a continuous output, is used in place of a classification decision tree. For a complete review the reader is referred to \cite{Criminisi2011}.

\subsubsection{ARF Learning:} One ARF is learned separately for each training image and dose pair. Learning is performed by constructing a binary tree, $T_{t,j}$, where each node branches to its left-or-right child node based on a learned decision rule. Each node predicts a PDF $P(d_{j,\pixel}|F_{*,j,\pixel})$ over possible dose values, through an empirical estimate (analogous to a histogram over the data samples processed by that node).

The root node begins with subset of data samples randomly chosen from the atlas image, $I_j(\pixel)$. We use the classical decision rule of $f\le t$ for observed feature $f$, and learned threshold $t$. The threshold at each node is learned by minimizing the least-squared-error of the predicted dose distribution in the resulting child nodes. For example, splitting the data so that $5$ Gy samples are on the left, and $40$ Gy on the right. The forest, $\mathbf{T}_j$, is a set of trees, each learned over a different bootstrapped sample from the training image, thus using a combination of weaker learns to build a stronger learner \cite{Breiman2001}.

\subsubsection{Inference (atlas-to-dose mapping):} Per-voxel inference for a novel image involves calculating the features, and then for a single ARF traversing every tree starting from the root, evaluating $f\le t$ for each node, branching accordingly, and then returning the observed dose estimate of $P_{*,j}(d_{a,\pixel}|F_{*,a,\pixel}=f)$ at the reached leaf. 

In canonical RF, each forest would instead be trained on a random subset of voxels from a random subset of images, and inference would thus be averaged over many images. A key observation is that this type of inference only explicitly considers all voxels independently. The features are taken over sets of voxels via convolution, but a joint observation/estimate of the dose is not made across the set of all observed voxel features $F_{*,a,*}$. However the dose at each voxel is heavily dependent on its neighborhood, for example being close to the target, i.e. the contextual information. In essence, though we have thus far modelled  $P(d_{a,\pixel}|F_{*,a,\pixel})$, we ultimately wish to model $P(d_{a,\pixel}|F_{*,a,*})$, i.e. dose given the contextual information about the image appearance as a whole. In what follows we use density estimation and atlas-selection learning to accomplish this.

\subsubsection{Density estimation:} In \cite{mcIntosh2016TMIAutoPlan} we introduced using density estimation at the leaf-nodes to estimate the probability of observing a set of features given a particular atlas. Specifically, in addition to a feature threshold and $P(d_{j,\pixel}|F_{*,j,\pixel})$ each leaf-node further contains a multivariate normal distribution over the observed features on the path from the root to the leaf: $\mathcal{N}(\mu_{t,l,j},\Sigma_{t,l,j})$ for tree $T_{t,j}$ from training image $j$. Means and co-variances are computed over the data samples at the respective leaf, i.e. $\mu_{t,l,j}$ is for leaf $l\in T_{t,j}$ from data in $I_j$. Further, define $\mathcal{N}(\mu_{t,l,a},\Sigma_{t,l,a})$ for some novel data being pushed through the ARF from image $I_{a}(\pixel)$.

Consider a trivial tree with only a root node and two children, split on the raw CT value, for example. The decision learning determines the optimal dose prediction feature for the sampled voxels, and the density estimates at the leaves produce a two-mode mixture model estimating the likelihood of observing any particular CT value, given the atlas image. Using the same tree, the data in a novel image specifies a new density distribution, and the differences between the learned and observed density distributions indicate the similarity between the novel image and the atlas image. They key is that the density distributions are estimated over the features relevant to dose prediction learned during tree optimization, and thus can ignore superfluous image information. \figref{fig::dictionaryLearning} displays an example, where all of the voxels in the image that are predicted by a leaf are encoded by a single colour corresponding to a breadth-first ordered labelling of tree leaves. This is not a visualization of the feature itself, but rather the voxels that have a strong response to the set of features that defines a particular leaf. Thus the colour correspondence of voxels in \figref{fig::dictionaryLearning}-d and -f illustrates the set of corresponding voxels for novel images that are used to calculate and compare $\mathcal{N}(\mu_{t,l,a},\Sigma_{t,l,a})$.  

The next step is to establish a distance between the learned leaf model (e.g. lung-like patch in training atlas) and the observed image data (e.g. slightly larger lung-like patch). The difference between the observed word at leaf $l$, and the learned word, is taken as the Bhattacharyya distance for multivariate normal distributions:
\begin{equation}
\label{eq::bhat}
\begin{array}{ll}
B(I_{j},I_{a}|l,T_{t,j}) = & \frac{1}{8}(\mu_{t,l,j}-\mu_{t,l,a})^{T}\Sigma^{-1}(\mu_{t,l,j}-\mu_{t,l,a})\\
&+\frac{1}{2}\ln\left(\frac{\det\Sigma}{\sqrt[2]{\det\Sigma_{t,l,j} \det\Sigma_{t,l,a}}}\right),
\end{array}
\end{equation}
\noindent where $\Sigma = 0.5\left(\Sigma_{t,l,a} + \Sigma_{t,l,j}\right)$, where the first term is related to the Mahalanobis distance, and the second term the differences between the co-variances \cite{bhattacharyya1946measure}.

\subsection{Atlas-Selection Learning}

With the individual ARFs and corresponding density estimates learned, in this learning phase we use cross-validation strictly within the training set to predict the dose with each ARF over all other training images, measure the accuracy of the predicted dose, and then use the density estimates to learn to predicted the observed accuracy. Finally, only the k ARFs with the highest predicted accuracy are used for a novel image. For example, for a novel CNS brain patient with a PTV  near the Brainstem, only ARFs from patients with similar PTVs near the Brainstem are selected and used for dose prediction. The atlas-selection step ensures that the dose-per-voxel is not only specified based on the features at a voxel, but based on the contextual information in the image (i.e. the observed features at other voxels as a group). The atlas-selection step enables us to better approximate $P(d_{a,\pixel}|F_{*,a,*})$ since each atlas inherently models the voxel dose interdependence (see \cite{mcIntosh2016TMIAutoPlan} for details).

With \eqnref{eq::bhat}, the distance between two images can be computed as the sum over all the leaves, which is in-turn a sum over the entire image. However, entire areas of the image may not be relevant to contextual dose prediction, and hence atlas-selection. For example, when planning RT for a right-sided atypical lung target, the inferior left lung of the patient can be any shape or size and it will have no impact on the dose distribution. Instead of assuming a linear combination, we perform regression using a second independent RF to compute the final distance (denoted as a pRF). We define $E(\tilde{d}_{a,*},d_{a,*}|\mathbf{T}_j)\in \mathcal{R}^+$ as the difference between the predicted dose distribution $\tilde{d}_{a,*}$ using ARF $\mathbf{T}_j$ and the clinical dose distribution $d_{a,*}$. In \cite{mcIntosh2016TMIAutoPlan} we focused on the Gamma metric \cite{low2003evaluation}, but in this work we combine it with a sum of the absolute difference between the DVHs in the clinical and predicted plans to bring both a spatial and a total distribution-oriented context to the metric.

Each pRF is then trained using $B(I_{j},I_{a}|l,\mathbf{T}_{t,j})$ for $l\in\mathbf{T}_{t,j}$ as input features, over  $a\in \{1...j-1,j+1,...M\}$ samples with  $E(\tilde{d}_{a,*},d_{a,*}|T_j)$ as the explanatory variable, i.e. we use leave-one-out cross-validation over the training data set. The testing data is kept completely separate. Using each leaf as a seperate feature enables the pRF to learn to ignore any leaf,  $l\in\mathbf{T}_{t,j}$, that represents an area of the image that is not relevant to atlas-selection. For example, a group of rectum-like patches can form an overall rectum shape, and the detection of that shape in the novel image implies a high predictive accuracy for the ARF. 

In summary, each ARF learns a feature-based set of descriptors for its training image, and is paired with a pRF that predicts the accuracy of the ARF for any input image as a function of the learned descriptors at the ARF's leaves. For computational speed, a sub-sampling rate of $1.5\%$ is used to select voxel from the image for the density estimation.

We now write our equation for Contextual Atlas Regression Forests as:

\begin{equation}
\label{eq::CRF}
P(d_{a,\pixel}|F_{*,a,*}) = \sum_{j=1..M}P_{*,j}(d_{a,\pixel}|F_{*,a,\pixel},\mathbf{T}_j)P(\mathbf{T}_j|F_{*,a,*}),
\end{equation}
\noindent where $P(\mathbf{T}_j|F_{*,a,*})$ determines the likelihood of a RF $\mathbf{T}_j$ given the observed features of the test image which is used as the weighting of the forest prediction. Intuitively, the closer the test image features are to the forest representation, the higher the weight. In this work we use an equal weighting among the $k=4$ ARFs with minimally predicted error, where 4 was learned through cross-validation on the training data in \cite{mcIntosh2016TMIAutoPlan}.

\subsection{Conditional Random Field Dose Estimation}

The final step is to infer the most likely spatial dose distribution from $P_{*,j}\left(d_{a,\pixel}|F_{*,a,\pixel}\right)$. In \cite{mcIntosh2016TMIAutoPlan} we proposed to use maximum-a-posteriori (MAP) estimation per voxel. This approach is fast, but assumes independence among voxels, and therefore does not consider the total distribution of dose during inference. The distribution of dose is not to be confused with the dose distribution, the former is analogous to a DVH. In this work we propose to use the atlas-patients to estimate a joint prior for the distribution of dose to the target and ROIs, and then query $P_{*,j}\left(d_{a,\pixel}|F_{*,a,\pixel}\right)$ to find the most likely spatial distribution of dose to each voxel that adhere to the prior. 

Incorporating an independence assumption and a uniform dose prior, our previous work estimated scalar dose from the predicted PDF using a MAP estimate of the form:

\begin{equation}
\label{eq::mapEstimateInd}
\tilde{d}_{a} =  \argmax_{d_{a,*}} \prod_{\pixel} P_{*,j}\left(d_{a,\pixel}|F_{*,a,\pixel}\right).
\end{equation}

\noindent A MAP estimate is used as opposed to an average, as is more typical in RF, to avoid degradation of the dose by a single outlier, e.g. $[1\; 1\; 1\; 1\; 0]$ across five trees should predict a dose of $1$, not $0.8$.

Conversely, in this work we define a dose prior $P\left(d_{a,*}\right|R(\pixel))$ where $R(\pixel)$ is a binary vector denoting the ROI class membership of $\pixel$, from the set of targets and OAR ROIs $\mathcal{C}$. Note that in contrast to a typical DVH, which is a cumulative prior distribution per ROI, we employ a joint prior and thus the prior is of dimensionality $|\mathcal{C}|$. The dose prior is computed as the average prior over the most similar atlases as determined by the pRFs, similar to \eqnref{eq::CRF}. 

Finding the most likely spatial assignment of dose-per-voxel according to $P_{*,j}\left(d_{a,\pixel}|F_{*,a,\pixel}\right)$ under the dose prior can be written as a CRF:

\begin{equation}
\label{eq::mapEstimatePrior}
\tilde{d}_{a} =  \argmax_{d_{a,*}} \prod_{\pixel} P_{*,j}\left(d_{a,\pixel}|F_{*,a,\pixel}\right)P\left(d_{a,*}|R(\pixel)\right),
\end{equation}

\noindent which written in this form is a non-convex continuous optimization problem. We solve by transforming to a binary linear program:

\begin{equation}
\label{eq::CRFSolution}
\tilde{d}_{a} =  \argmax_{d_{a,*}} \sum_{\pixel} \sum_{g}log\left(P_{*,j}\left(d_{a,\pixel,g}|F_{*,a,\pixel}\right)\right)b_{\pixel,g},
\end{equation}

\noindent subject to the constraints:
\begin{equation}
\begin{array}{ccc}
\sum\limits_g b_{\pixel,g} = 1 & \mathnormal{for} & \pixel \in \Omega\\
b_{\pixel,g}\geq 0 & \mathnormal{for} & \pixel \in \Omega, g\in G,\\
\sum\limits_{\pixel\in R} b_{\pixel,g} \leq |R|P\left(d_{a,*,g}\right|R(\pixel)) & \mathnormal{for} & R \in \mathcal{C},g \in G\\
\end{array}
\end{equation}
\noindent where $g$ is introduced to discretize the dose into $G$ bins, and $b_{\pixel}$ is a binary indicator variable. The first two constraints ensure that the binary indicator specifies at most one dose value-per-voxel. The third constraint enforces that the dose satisfies the required prior, with one such prior per ROI intersection sub-type (e.g. a voxel being just lung, or target and lung). 

\newcommand{\figWidth}{0.13\linewidth}
\newcommand{\figWidthC}{0.168\linewidth}
\begin{figure*}[!t]
\footnotesize
	\begin{center}
		\begin{tabular}{cccccc}
			\includegraphics[width=\figWidth,clip=]{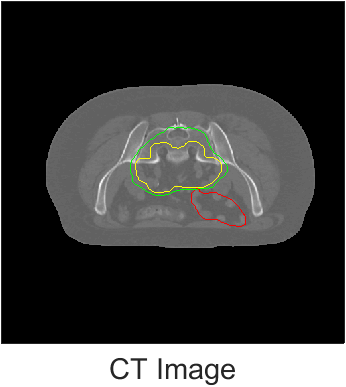}&\includegraphics[width=\figWidthC,clip=]{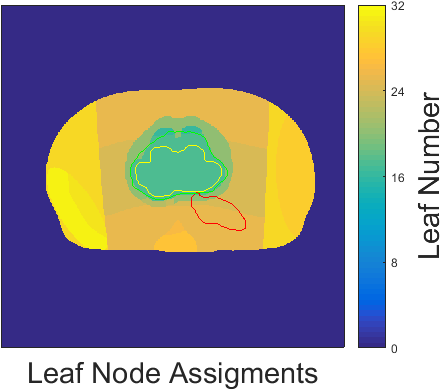}&\includegraphics[width=\figWidth,clip=]{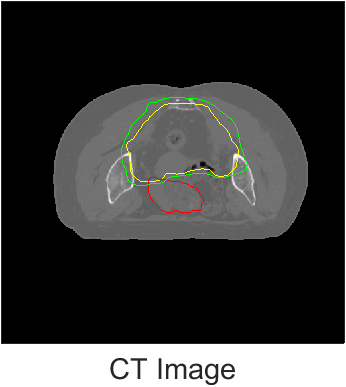}&\includegraphics[width=\figWidthC,clip=]{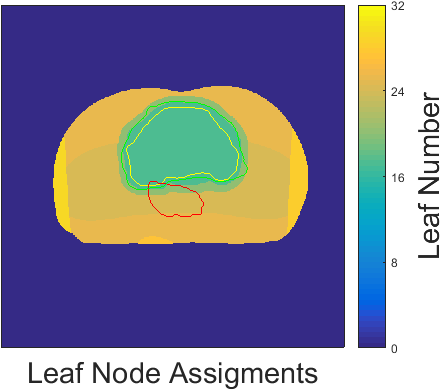}&\includegraphics[width=\figWidth,clip=]{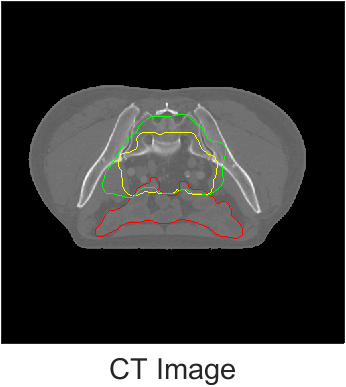}&\includegraphics[width=\figWidthC,clip=]{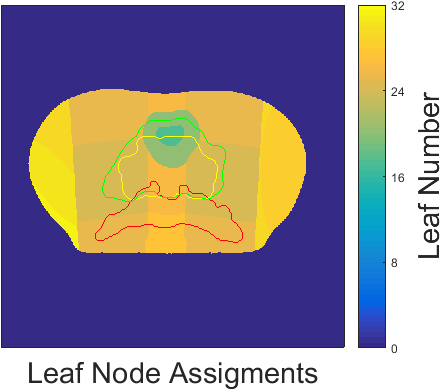}\\
            \includegraphics[width=\figWidth,clip=]{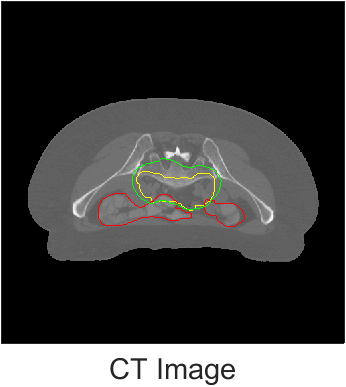}&\includegraphics[width=\figWidthC,clip=]{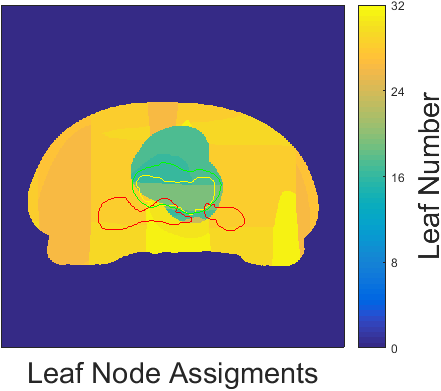}&\includegraphics[width=\figWidth,clip=]{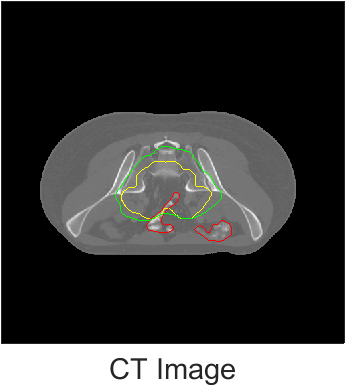}&\includegraphics[width=\figWidthC,clip=]{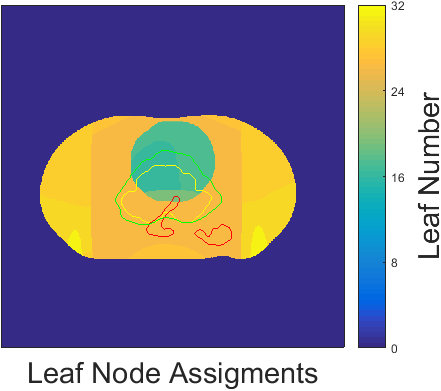}&\includegraphics[width=\figWidth,clip=]{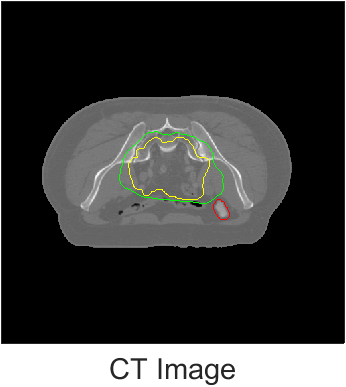}&\includegraphics[width=\figWidthC,clip=]{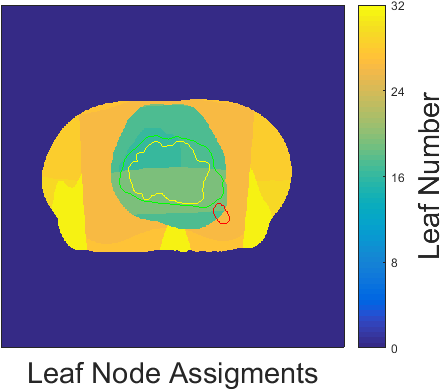}\\
    			\centre{2}{Atlas} & \centre{2}{Neighbouring} & \centre{2}{Distant}\\
			(a) & (b) & (c) & (d) & (e) & (f)\\

		\end{tabular}
	\end{center}
	\caption[Dictionary learning]{(Colour Figure) Atlas Regression Forest dictionary learning for rectum. Atlas-selection automatically becomes a function of small bowel proximity and intersection with the target. Irrelevant variation, e.g. external shape is ignored. (a) CT image slices of various atlas patients in each row. (b) Corresponding labelled atlas image slices where colour indicates the leaf number under a breadth-first labelling of tree leaves of the respective Atlas Regression Forest. (c-d, e-f) pairs of novel images that have high/low dose prediction accuracy under the atlas in a given row, respectively. Dose at $90\%$ of prescription is contoured on each image slice in green, the clinical target in yellow, and the small bowel in red for reference. Notice the similarity between (a) and (c) in comparison to (a) and (e). Also notice the difference between the two rows, automatically learning that target and small bowel proximity is important for global dose geometry.} \label{fig::dictionaryLearning}
\end{figure*}

\section{Results}
\label{sec::results}
\par We divide the data into independent training and testing sets (\tabref{tb::dataSummary}). Breast, and prostate sites were used for algorithm development and parameter tuning using cross-validation on the training set in \cite{mcIntosh2016TMIAutoPlan}, and those same parameter values are used here. The lung, rectum, and brain data use the same parameters without re-tuning.
 
We evaluate variants of both the atlas-selection algorithm and the atlas-dose to novel image mapping. For atlas-selection we examine using: no atlas-selection, the overlap-volume-histogram (OVH) distance between patients \cite{Kazhdan2009}, and the proposed DVH-Gamma hybrid. For mapping the dose onto the novel image we experiment with: deformable image-based atlas-registration (DAR), and our ARFs. For DAR, the CT images are first registered with translation only since all patients are in the same pose for treatment. Deformable registration is carried out between the CT images with the default parameters in \cite{yang2011technical}. Registration is performed in 4 multi-resolution phases at scales of: 1/32, 1/16, 1/8, and 1/4. The dose from the atlas image is warped onto the novel image via the same transformations (translation, then deformable). We also include a generic OVH method where the DVH is taken directly from the nearest patients under the OVH metric.

Approximate upper-bounds for ARFs and DAR are computed by testing all training atlases on all test images. In order to compute a fast upper-bound approximation the images were sub-sampled to $1/4$ size in all dimensions. Negligible differences were observed at full image resolution. However, note that the ARF-CRF can outperform the upper-bound, as the bound focuses on atlas-selection and is computed without the aid of a CRF.

\begin{figure*}[!t]
	\begin{center}
			\includegraphics[width=0.90\linewidth,clip=]{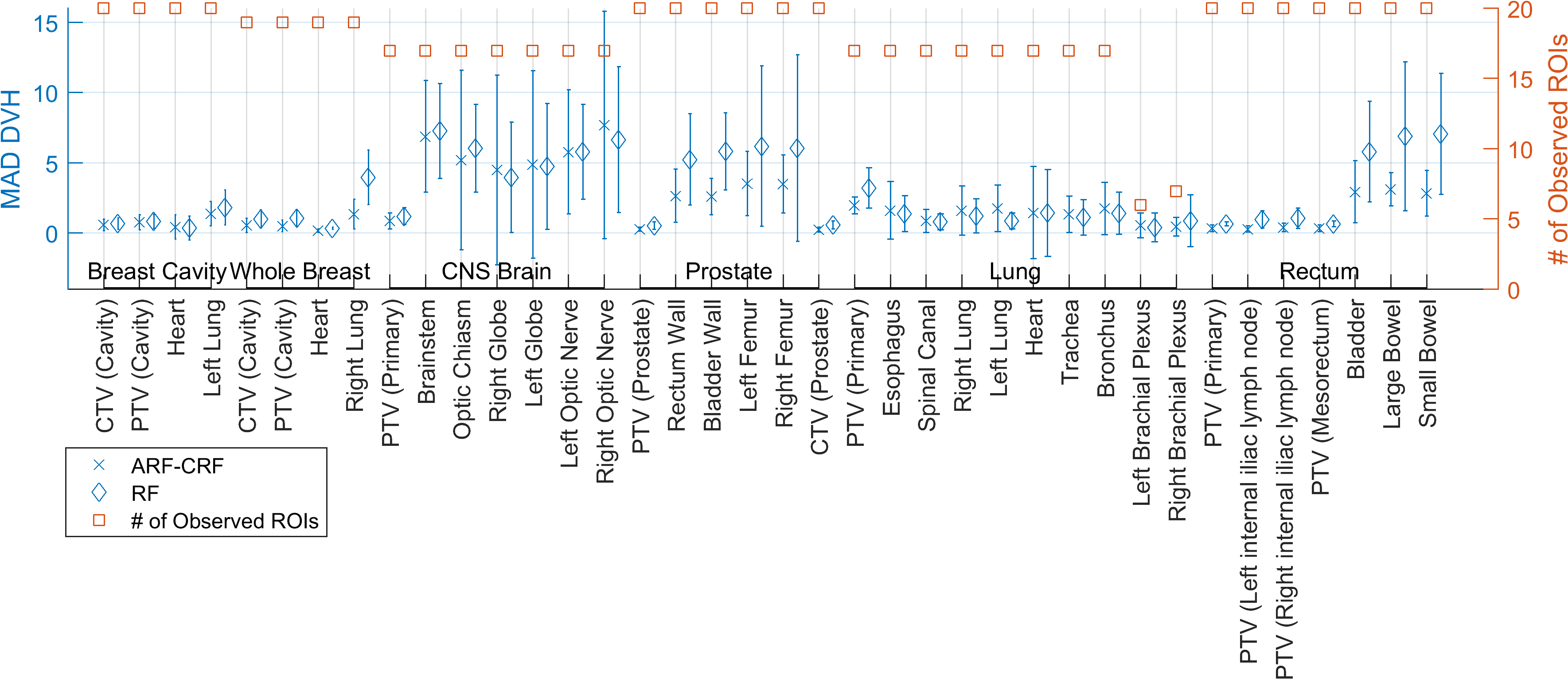}
	\end{center}
	\caption[Dictionary learning]{(Colour Figure) Distribution of Mean Average Difference (MAD)-DVH errors across patients for all sites and ROIs with error bars representing one standard deviation.} \label{fig::dvhErrorsAllROIs}
\end{figure*}

Quantitative results are presented in \tabref{tb::IMRT}. The error is measured as the mean absolute difference (MAD) between the clinical DVH and the DVH calculated from the predicted dose distribution averaged over the set of ROIs, and the tested patients. The minimum value is $0\%$, and the maximum is $100\%$. Standard deviation is calculated between patients. The $95\%$ confidence interval (CI) is calculated from a paired two-tailed t-test with the MAD-DVH as the dependent variable, comparing the proposed ARF-CRF with the other methods. We corrected for multiple-comparison testing with Bonferroni correction carried out over the 6 tests for each site. The upper bounds are excluded from this result since they are not actually realizable for a novel patient. \figref{fig::dvhErrorsAllROIs} displays plots of the MAD-DVH error for each ROI in each site individually, including the number of observed instances of the ROI. ROIs were not contoured in plans where they appear outside the field and/or receive little to no dose. \figref{fig::resultSummary} displays example dose overlays for both clinical and automated plans under the proposed ARF-CRF pipeline. 

\begin{table}[!ht]
\caption[Summary of results.]{Mean Average Difference (MAD) between automated and clinical plan Dose Volume Histograms for all Regions of Interest (ROIs).}
\label{tb::IMRT}
\footnotesize
\begin{tabular*}{\linewidth}{@{}llllllll}
\br
&&\centre{2}{Breast Cavity}&\centre{2}{Whole Breast}&\centre{2}{CNS Brain}\\
\textbf{Map} & \textbf{Atlas} & \textbf{MAD-} & \textbf{CI}$^b$& \textbf{MAD-} &  \textbf{CI}& \textbf{MAD-} &  \textbf{CI}\\
\textbf{} & \textbf{} & \textbf{DVH}$^a$ & \textbf{}& \textbf{DVH} &  \textbf{}& \textbf{DVH} &  \textbf{}\\
\mr
u.b. ARF & DVH$+\Gamma$ & 0.87$\pm$0.46 & - & 1.23$\pm$0.57 &  - & 3.65$\pm$1.20 &  - \\ 
ARF & DVH$+\Gamma$ & 0.96$\pm$0.53 &  [-0.4,0.0] & 1.19$\pm$0.67 &  [-0.8,-0.3] & 5.35$\pm$2.26 &  [-1.2,0.7] \\ 
\underline{ARF-CRF} & DVH$+\Gamma$ & 0.79$\pm$0.39 &  - & 0.64$\pm$0.43 &  - & 5.11$\pm$2.74 &  - \\ 
Generic & OVH & 1.12$\pm$0.53 &  [-0.6,-0.1] & 0.88$\pm$0.29 &  [-0.4,-0.1] & 8.01$\pm$3.20 &  [-5.1,-0.7] \\ 
ARF & OVH & 1.36$\pm$0.56 &  [-0.9,-0.3] & 2.01$\pm$2.07 &  [-2.6,-0.1] & 6.78$\pm$3.01 &  [-4.5,1.2] \\ 
RF & None & 0.92$\pm$0.51 &  [-0.3,0.1] & 1.59$\pm$0.66 &  [-1.2,-0.6] & 5.09$\pm$2.09 &  [-1.4,1.5] \\ 
u.b. DAR & DVH$+\Gamma$ & 7.52$\pm$6.55 &  - & 1.22$\pm$0.40 &  - & 4.98$\pm$3.44 &  - \\ 
DAR & DVH$+\Gamma$ & 9.54$\pm$11.79 &  [-16.4,-1.1] & 2.46$\pm$1.43 &  [-2.6,-1.0] & 8.19$\pm$4.54 &  [-6.3,0.1] \\ 
DAR & OVH & 19.93$\pm$16.26 &  [-29.7,-8.6] & 5.34$\pm$9.76 &  [-11.1,1.7] & 9.32$\pm$4.34 &  [-7.8,-0.6] \\ 
\mr
&&\centre{2}{Prostate}&\centre{2}{Lung}&\centre{2}{Rectum}\\
\mr
u.b. ARF & DVH$+\Gamma$ & 2.53$\pm$0.67 &  - & 1.33$\pm$0.93 &  - & 2.40$\pm$1.01 &  - \\ 
ARF & DVH$+\Gamma$ & 3.21$\pm$1.16 &  [-1.9,-0.3] & 1.62$\pm$0.85 &  [-0.6,0.0] & 3.09$\pm$1.26 &  [-2.6,-0.7] \\ 
\underline{ARF-CRF} & DVH$+\Gamma$ & 2.13$\pm$0.75 &  - & 1.33$\pm$0.74 &  - & 1.46$\pm$0.54 &  - \\ 
Generic & OVH & 2.53$\pm$1.21 &  [-1.1,0.3] & 2.68$\pm$1.19 &  [-1.9,-0.8] & 4.34$\pm$1.21 &  [-3.6,-2.2] \\ 
ARF & OVH & 6.38$\pm$3.96 &  [-6.9,-1.6] & 3.22$\pm$2.49 &  [-3.4,-0.4] & 4.42$\pm$2.72 &  [-4.8,-1.1] \\ 
RF & None & 4.07$\pm$2.39 &  [-3.7,-0.2] & 1.27$\pm$0.69 &  [-0.3,0.4] & 3.29$\pm$1.62 &  [-3.0,-0.7] \\ 
u.b. DAR & DVH$+\Gamma$ & 2.74$\pm$0.76 &  - & 7.96$\pm$3.31 &  - & 2.92$\pm$0.81 &  - \\ 
DAR & DVH$+\Gamma$ & 4.92$\pm$1.94 &  [-4.0,-1.6] & 10.53$\pm$4.13 &  [-11.8,-6.6] & 5.45$\pm$1.84 &  [-5.2,-2.7] \\ 
DAR & OVH & 5.01$\pm$2.26 &  [-4.5,-1.3] & 13.84$\pm$3.81 &  [-14.9,-10.1] & 6.18$\pm$2.57 &  [-6.3,-3.1] \\ 
\end{tabular*}
\newline
\begin{tabular*}{\linewidth}{@{}l@{\extracolsep{\fill}}lll}
\mr
Notes:&&&\\
\textbf{Abbreviation} & \textbf{Definition} &\textbf{Abbreviation} & \textbf{Definition}\\
\mr
u.b. & Upper Bound (Approximate) & ARF & Atlas Regression Forest\\
CRF & Condition Random Field & RF & Regression Forest\\
DAR & Deformable Atlas Registration & DVH & Dose Volume Histogram\\
OVH & Overlap Volume Histogram & $\Gamma$ & Gamma Analysis Metric\\
\br
\multicolumn{4}{l}{$^a$- Mean $\pm$ standard deviation calculated across patients}\\
\multicolumn{4}{l}{$^b$- $95\%$ Confidence Interval for mean difference in MAD-DVH in comparison to ARF-CRF}\\
\end{tabular*}
\end{table}

\newcommand{\DosefigWidth}{0.13\linewidth}
\begin{figure*}[ht]
	\begin{center}
     \footnotesize
		\begin{tabular*}{\linewidth}{@{}ccccccl}
			\includegraphics[width=\DosefigWidth,clip=]{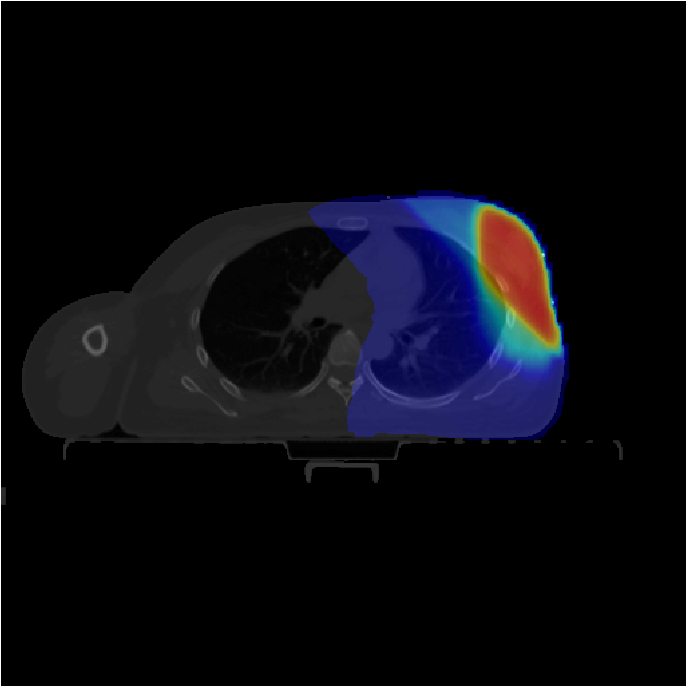}&
            \includegraphics[width=\DosefigWidth,clip=]{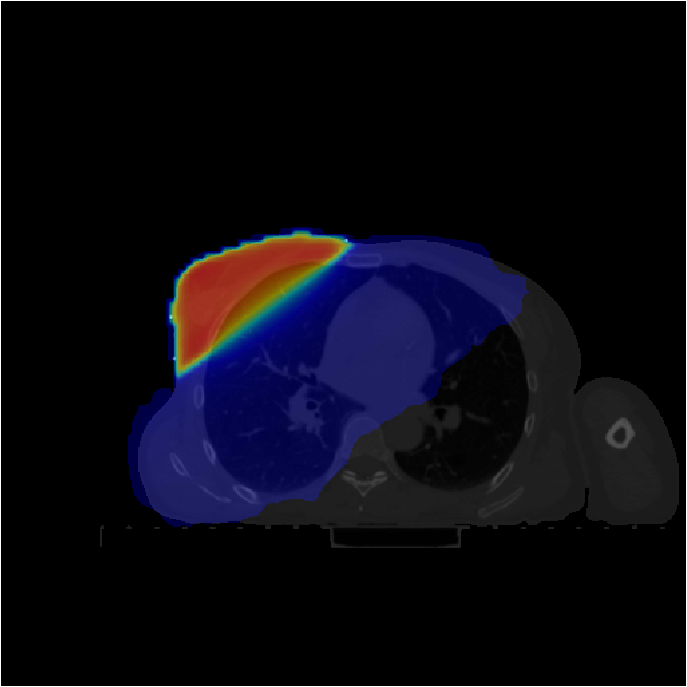}&
            \includegraphics[width=\DosefigWidth,clip=]{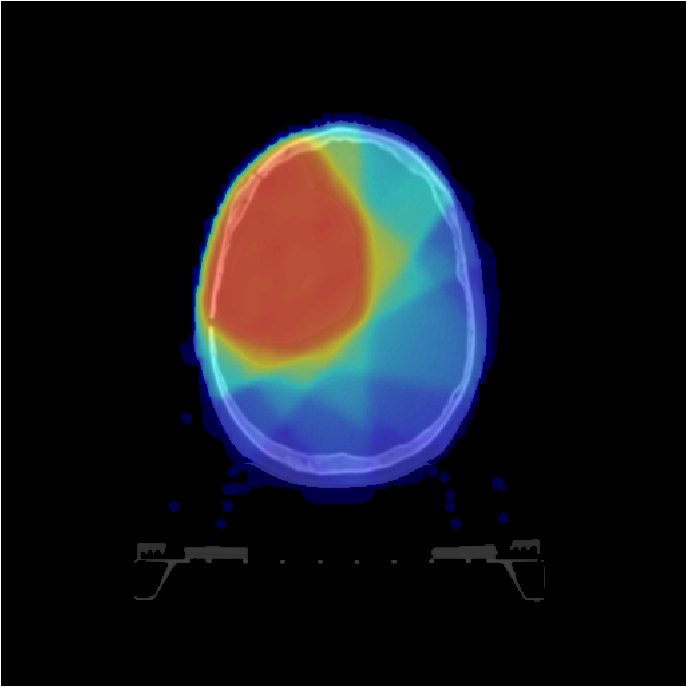}&
            \includegraphics[width=\DosefigWidth,clip=]{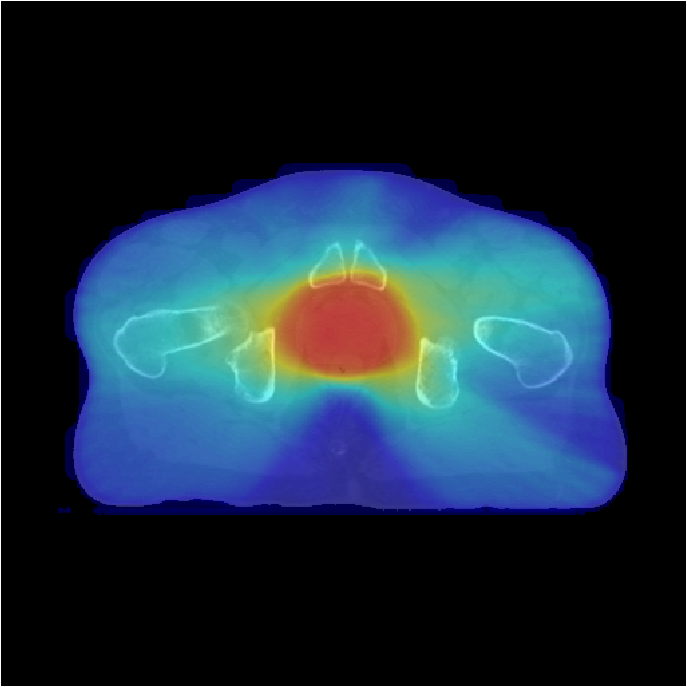}&
            \includegraphics[width=\DosefigWidth,clip=]{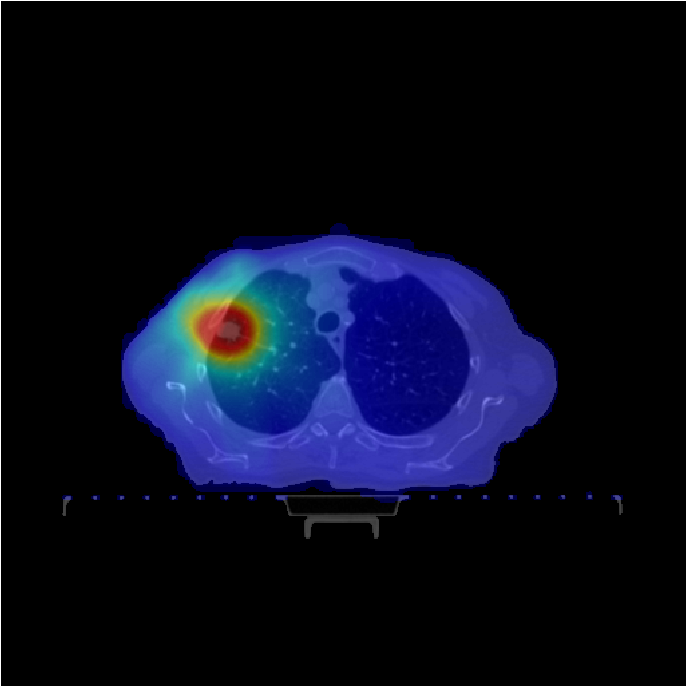}&
            \includegraphics[width=\DosefigWidth,clip=]{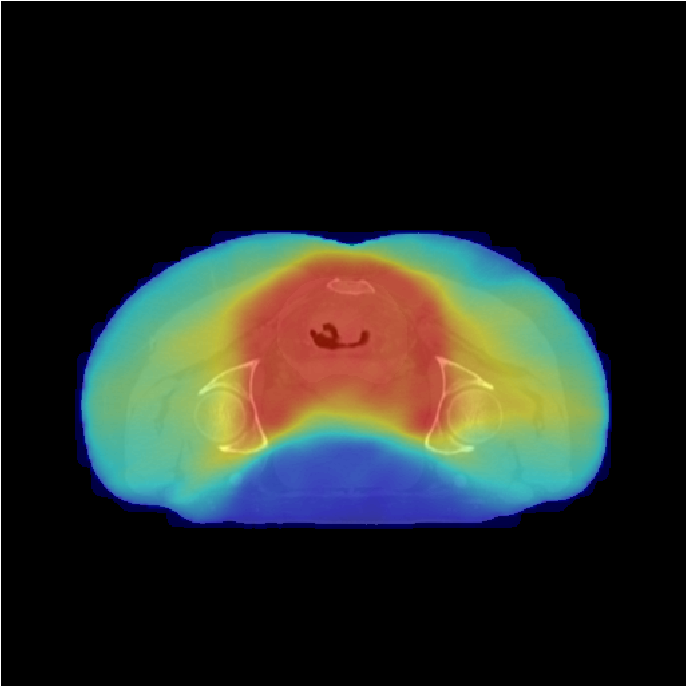}&\multirow{2}{*}[40pt]{ \hspace{-10pt}\includegraphics[width=0.06\linewidth,clip=]{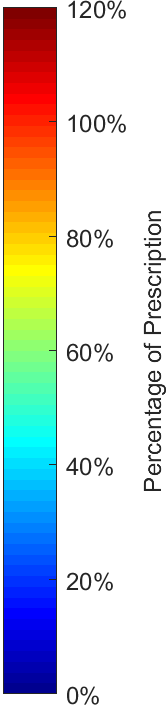}}\\
			\includegraphics[width=\DosefigWidth,clip=]{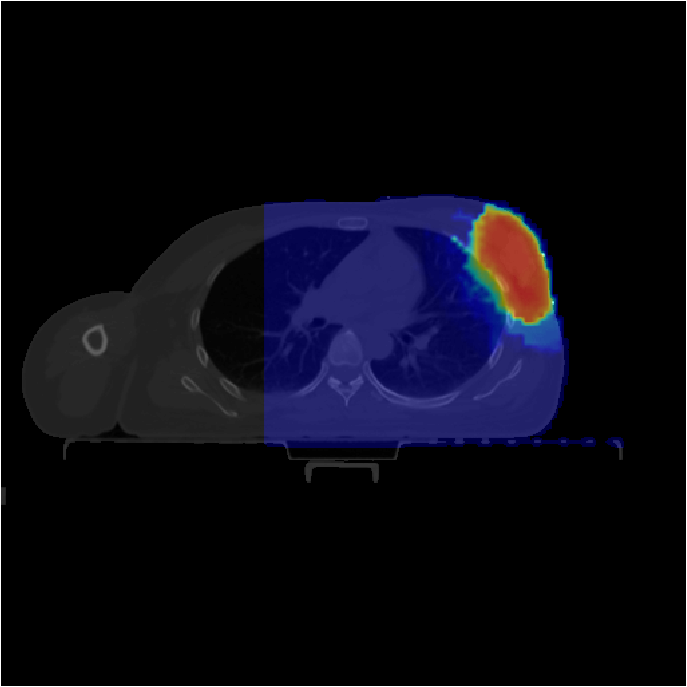}&
            \includegraphics[width=\DosefigWidth,clip=]{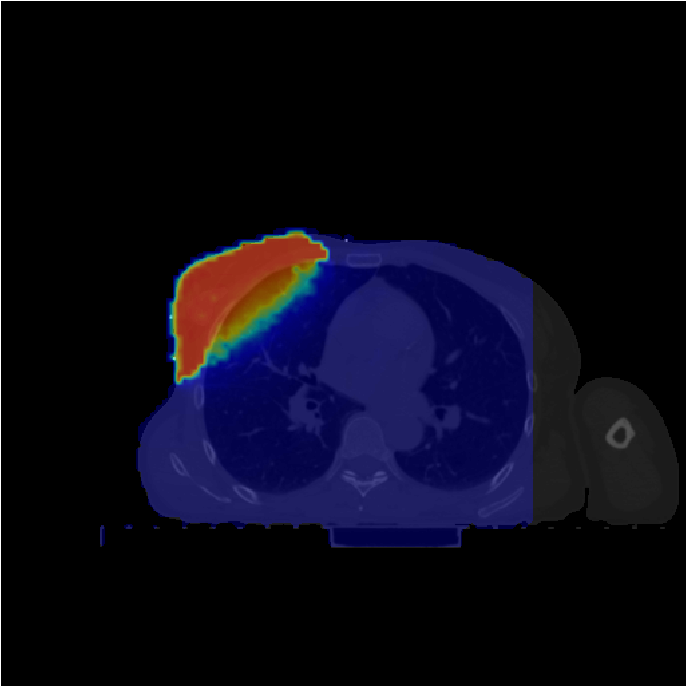}&
            \includegraphics[width=\DosefigWidth,clip=]{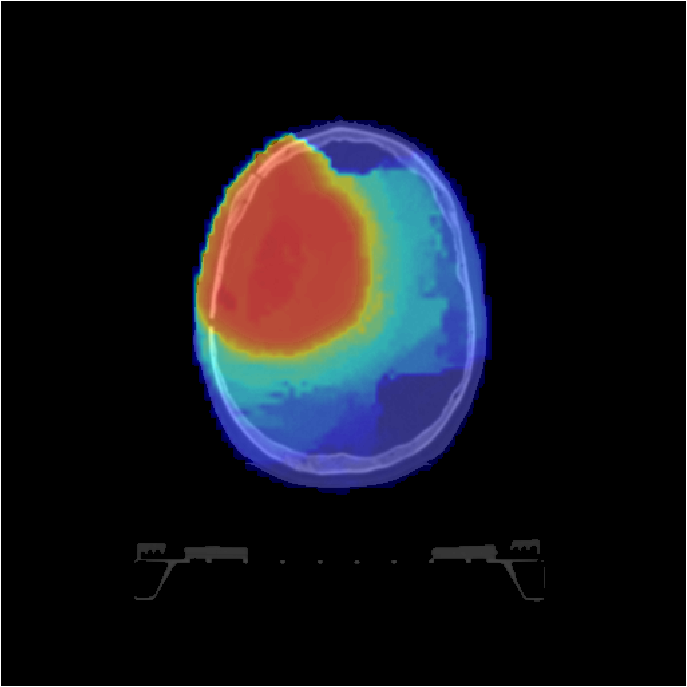}&
            \includegraphics[width=\DosefigWidth,clip=]{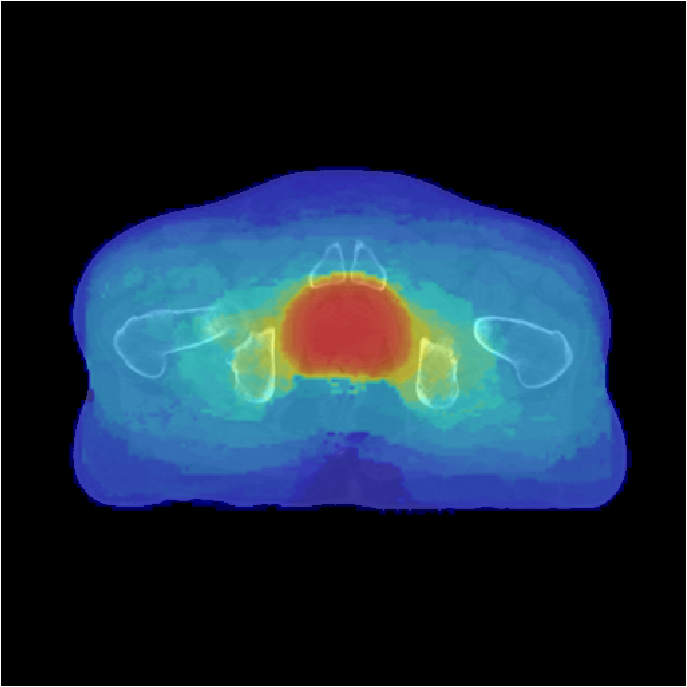}&
            \includegraphics[width=\DosefigWidth,clip=]{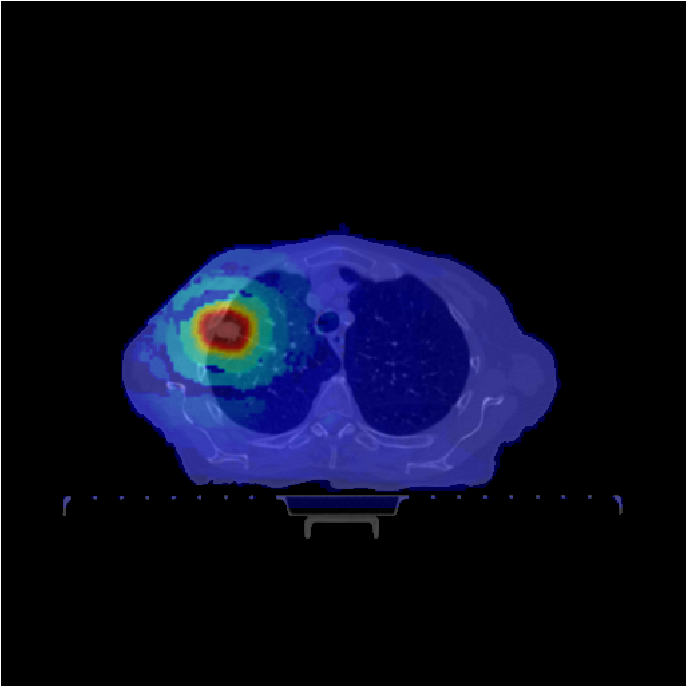}&
            \includegraphics[width=\DosefigWidth,clip=]{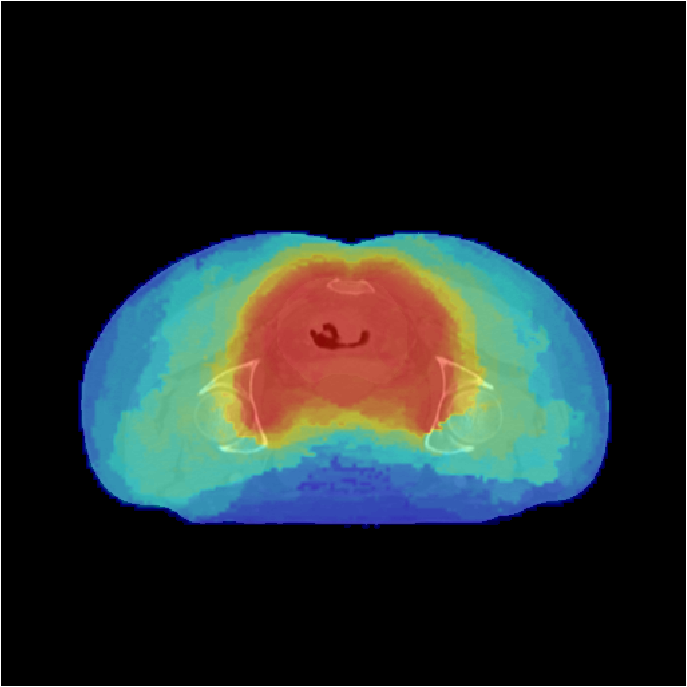}&\\
            Breast Cavity & Whole Breast & CNS Brain & Prostate & Lung & Rectum\\
		\end{tabular*}
	\end{center}
	\caption[]{(Colour Figure) Visualizations of dose overlays for one example of each site. Dose is normalized to a percentage of prescription for each case (with red indicating $100\%$ of prescription). (Top) Clinical dose. (Bottom) ARF-CRF predicted dose distribution.} \label{fig::resultSummary}
\end{figure*}

The algorithm is reasonably fast, with a total run-time including feature calculation, atlas-selection, and CRF optimization of approximately 15-minutes for a novel image volume. For speed the CRF is solved on the dose-grid, as opposed to the image grid. Novel sample prediction is carried out independently across trees in the ARFs and thus performance scales well with the system specifications.

\section{Discussion and Related Work}
\label{sec::discussion}
\par 
The current paper extends the voxel-based dose prediction method previously presented \cite{mcIntosh2016TMIAutoPlan}, by improving the features for altas selection to include DVH estimation accuracy, and incorporating CRFs to optimize a joint prior for dose prediction. We have also provided a more comprehensive clinical evaluation for six treatment sites and two treatment modalities and provided evaluation based on DVHs for targets and relevant OARs. The method presented requires only a limited number of ROIs \figref{fig::dvhErrorsAllROIs} (e.g. four for Breast Cavity and ten for Lung) to generate a dose distribution. By modeling the image appearance within ROIs and the appearance and geometrical relationships of non-delineated structures in the images we can better patient similarity than approaches using only ROI data. In this way, we can generate a SDO automatically, that does not require any dose objectives to be specified explicitly; the SDO captures all the necessary information to drive dose prediction.
\par
There are a number of related works in automated planning \cite{kazhdan2009shape,appenzoller2012predicting,wu2013using,yang2013overlap,shiraishi2015knowledge,shiraishi2016knowledge}. Most of the work in this area focuses on using high level shape descriptors of contoured structures to retrieve relevant DVHs from a database, and then infer corresponding dose volume objectives \cite{appenzoller2012predicting,wu2013using,yang2013overlap}. After specifying the dose-volume objectives, planning proceeds as normal. However, recent independent work by Shiraishi and Moore \cite{shiraishi2016knowledge}, and simultaneously our group \cite{mcIntosh2016TMIAutoPlan} has introduced the notion of predicting the dose on a voxel-by-voxel basis. Shiraishi and Moore used groups of previous patient image-dose pairs to train two dose prediction models: one inside the target and one outside. They performed validation of the predicted dose distributions for 12 prostate and 23 stereotactic radiosurgery plans, demonstrating improved OAR sparing in comparison to DVH models \cite{appenzoller2012predicting,shiraishi2015knowledge}.

However, in preface to discussion of voxel-based dose prediction methods it is important to note that the resulting dose distributions from \cite{mcIntosh2016TMIAutoPlan,shiraishi2016knowledge} and those presented here are not necessarily achievable, and thus a stricter evaluation criteria than improved sparing is warranted. Although \cite{shiraishi2016knowledge} suggested predicted dose distributions are deliverable since they are generated based on deliverable plans, three-dimensional dose distributions do not form a vector space because dose is purely additive; a negative weighted linear combination of two dose distributions does not necessarily create an achievable dose distribution. In addition the models perform inference per-voxel, and thus are incapable of ensuring spatial consistency across voxels. Therefore, predicting increased sparing or coverage may not actually be achievable when plans are generated to emulate the predicted dose. That said, this caveat should not detract from work in this direction. As the limit in the absolute difference between the predicted dose distribution, and the clinical dose distribution tends to zero, the predicted dose will by definition be achievable. Hence, in contrast to signed deviation, absolute value deviation penalizes any discrepancy and the more closely the predicted and clinical dose distribution match, the greater the predicted dose can readily be emulated. Thus evaluation of these model should focus on the absolute value deviations from clinical dose, as is done here. 

Similar to \cite{shiraishi2016knowledge} we do observe trends in increased sparing in cases where the absolute error is larger, and it will be very interesting to examine inverse-planning on a large scale with inverse-planning and SDOs to learn if this is achievable. However, the best mechanism through which to do so remains open, and is a separate question from how best to infer the proposed spatial distribution itself.

\par Turning to our results, we compared a variety of different atlas-selection and atlas-novel image dose mapping algorithms. Averaged across all treatment sites, the best performer was the proposed CRF-ARF method ($1.91$) followed by our previous ARF method without CRF \cite{mcIntosh2016TMIAutoPlan} ($2.57$) and RF without atlas-selection ($2.71$). Examining individual sites and comparing the upper-bound DVH$+\Gamma$ atlas-selection to RF without atlas-selection indicates that atlas-selection is impactful for Whole Breast, CNS Brain, Prostate and Rectum, but negligible for Breast Cavity and Lung. This result is logical given that Breast Cavity and Lung have high conformity with little sacrifice of target dose due to OARs, and hence the dose is a consistent function of the distance to the target. While atlas-selection should help in CNS brain ($3.65$ vs $5.09$), in practice that is where we observed the largest performance gap between the ARF and its upper-bound, leading to a negligible improvement over the RF without atlas-selection. This indicates that future feature development could improve atlas-selection, and thus algorithm performance for this treatment site. Another potential improvement for this site is to calculate new features for OARs to help with atlas-selection. It is also feasible that there are many clinically acceptable plans with varying DVHs created by small perturbations in the IMRT beam geometry, thereby confounding atlas-selection.

The proposed ARF-CRF algorithm on average out-performed the ARF algorithm for all sites, but lacked statistical significance indicated by the CI in Breast Cavity, Lung, and Brain. We believe the reasoning is the same as above. If atlas-selection is less impactful (or realizable) than inferring the dose-prior, then adhering to it is less impactful as well. The CIs are mostly negative, and so a larger evaluation set might also improve significance. Interestingly, examining the individual ROIs in \figref{fig::dvhErrorsAllROIs} we note that atlas-selection always improves target performance. Comparing with and without atlas-selection in Lung RT we observed increased accuracy on the targets, but increased error on the lung dose which is a potential for increased sparing.

DAR is the overall worst performer ($9.93$), with reasonable Prostate and Whole Breast performance but the worst for Breast Cavity and Lung. Prostate has a very fixed target geometry, where as Lung and Breast Cavity have targets that are free to translate around within a larger region of the patient. Large translation of only the target position between an atlas and a novel patient force tearing in the deformation field, which is discouraged and resisted by DAR methods in-place of smoothing image warps. In contrast, the RF approaches can learn to predict dose as a function of voxel-to-target distance which easily adapts to moving the target a few inches laterally.

The generic OVH method serves as mostly a baseline, since it directly infers a DVH instead of a spatial dose distribution. It is interesting that the proposed atlas-selection is able to outperform OVH without having access to the OAR contours. We believe this can be attributed to the automatic sub-segmentation of the image into crucial regions (\figref{fig::dictionaryLearning}), which given the voxel-distance-target feature already included can implicitly calculate an OVH-like metric. When the ARFs are used for dose mapping, DVH$+\Gamma$ enables the addition of spatial dose information and thus improved performance over OVH, as it was not designed for this purpose.

Further, the ARF-CRFs demonstrate the increased ability of the joint distribution dose prior, over inferring a set of uni-variate dose priors (DVHs). In \cite{boutilier2016sample} they demonstrated that such a method needs a very large number of atlases. This is a result of intersections between ROIs creating high degrees of variability in the DVH, a problem not suffered by the joint distribution. For example, changing the size of the target by 10\% in Lung changes the lung DVH, but not the joint distribution. 

Lastly, in \cite{mcIntosh2016TMIAutoPlan} we observed that the number of needed training atlas was a function of training site complexity and variability. We found similar results in this study, for example, that atlas-selection was not crucial for Lung, and therefore requires fewer training images.

The method described can be extended and applied to include fluence optimization \cite{shepard1999optimizing} in order to generate deliverable dose distributions. Although not presented in the current paper, the method can be readily applied to other treatment modalities, such as particle therapy, as the method estimates dose directly from planning images without any planning information being used. Therefore we can learn how dose distributions are shaped. The method can also provide estimates of the beam geometry from the patient atlases selected for doing the dose distribution prediction. In addition, there are clear applications for applying the methodology to achieve rapid planning to facilitate adaptive planning strategies.

\section{Conclusions}
\label{sec::conclusions}
This paper describes a methodology for the rapid estimation of the patient dose distribution without requiring any dose-volume objective specification and requiring only the target and limited OARs as inputs. We have developed a highly general voxel-based dose prediction method incorporating multi-patient atlas training that has applicability across a range of clinical sites and treatment modalities and techniques. We examined variants of our voxel-based dose prediction method with and without atlas-selection, and show that for sites in which there is variability in target coverage and dose distribution shaping between patients, atlas-selection provided substantially better dose prediction results. Whereas sites which have high conformity and minimal shaping, atlas selection was less important. Based on these results, our future work is to examine our pipeline in conjunction with inverse-planning technologies to determine if atlas-selection can improve OAR sparing. We can apply the presented methodology as a new automated planning strategy or a method that can be readily integrated into existing optimization-based approaches.
\bibliographystyle{dcu}
\bibliography{autoPlan}

\end{document}